\documentclass{article}
\usepackage{graphicx}
\usepackage[numbers,round]{natbib}
\begin{document}
\bibliographystyle{plain} 
\title{Activation of Graphenic Carbon due to Substitutional Doping by Nitrogen: 
Mechanistic Understanding from First-principles}
\author{Joydeep Bhattacharjee\\
\small{\textit{School of Physical Sciences}}\\
\small{\textit{National Institute of Science Education and Research}}\\
\small{\textit{IOP campus, Sachivalay Marg, Bhubaneswar-751005, India}}}
\maketitle
\begin{abstract}
Nitrogen doped graphene and carbon nanotubes are popularly in focus 
as metal-free electro-catalysts for oxygen reduction reactions (ORR) central to fuel-cells. 
N doped CNTs have been also reported to chemisorb mutually, promising a route to their robust
pre-determined assembly into devices and mechanical reinforcements. 
We propose from first-principles a common mechanistic understanding of these two aspects 
pointing further to a generic chemical activation of carbon atoms due to substitution by nitrogen in 
experimentally observed configurations. Wannier-function based orbital resolved
study of mechanisms suggests increase in C-N bond-orders in attempt to retain $\pi$-conjugation
among carbon atoms, causing  mechanical stress and loss of charge neutrality of nitrogen and  
carbon atoms, which remedially facilitate chemical activation of N coordinated C atoms, 
enhancing sharply with increasing coordination to N and proximity to zigzag edges.
Activated C atoms facilitate covalent adsorption of radicals in general, 
diradicals like O$_2$ relevant to ORR, and also other similarly activated C atoms leading to 
self-assembly of graphenic nano-structures, while remaining inert to ordinary graphenic C atoms. 
\end{abstract}

\baselineskip 14pt
Disruption of $\pi$-conjugation due to substitution by $p-$ or $n-$ type dopants in graphene 
causes localization of 2$p_z$ electrons, which are a rich source of exotic physical 
as well as chemical functionalities unavialable in undoped graphene. 
Accordingly, boron (B) and nitrogen (N) doped graphene (GF) and carbon nanotubes(CNT) have been under rigorous 
scrutiny during the last decade or so, resulting into a 
great multitude of proposals for novel applications ranging from nano-electronics to catalysis. 
Electro-catalytic functionality of doped GFs and CNTs have been under investigation
\cite{yu2010,wang2012,ruitaro2012} for more than a decade, primarily in pursuit of an 
efficient metal-free catalyst to replace the expensive platinum based ones used in fuel-cells so far. 
In particular, N doped CNTs and 
GFs\cite{gong2009,qu2010,jafri2010,sheng2011,li2012,sharifi2012,cheng2012,zhao2013,vikkisk2014}, 
and more recently their composites\cite{chung2013,chen2013,ratso2014,tian2014,higgins2014}, 
have been explored exhaustively in the  last few years as catalysts for  oxygen reduction reactions (ORR)
promisimg removal of platinum from the cathode in fuel-cells.
On the other hand, N doping has also been suggested\cite{payne} to induce covalent mutual adsorption 
among CNTs through C-C cross-links, as has indeed been reported\cite{wang2011}, promising a new route to their 
controlled assembly into devices and stronger composites, which in itself is an important open problem.
In the present work we show mechanistically that indeed these two aspects are just different facets of a 
more inclusive general scenario of chemical activation of C atoms arising with experimentally observed 
configurations of substitutional doping by N in $sp^2$ hybridized carbon (C) based 
graphenic nano-structures (GN).

First principles studies\cite{hu2010,kim2011,zhang2011,yu2011,yan2012} of ORR catalysis so far, primarily 
of isolated mono-substitution by N, suggest adsorption of atomic, molecular as well as active anionic 
oxygen (O) on C atoms in the vicinity of quaternary N atoms, presumably driven by the effective positive charge of 
C atoms due to electron extraction by their N neighbours, which however does not explain the observed 
mutual adsorption of N doped CNTs. 
Interestingly, recent direct observations \cite{zhao2011,ruitaro2012-2} suggest substitution by N
in GF to occur predominantly at next-nearest neighbouring (N-Nn) sites, which hinders one of the three 
degenerate $\pi$-conjugation configurations more than the other two, unlike in case of sparse
mono-substitution where all three of them are equally hindered. This implored us to ask whether 
multiple substitution at N-Nn sites can open up a new class of mechanisms supporting ORR catalysis 
as well as GN-GN covalent adsorption on the same footing. 
Additionally, as the N-Ns sites belong to one of the two sub-lattices of graphene, which 
support inter-sub-lattice spin-separation upon physical or chemical disparity between the two sub-lattices, 
the resultant magnetism needs to be linked to other functionalities arising out of substitution at N-Nn sites.
Notably, inter-sublattice spin-separation, which manifests itself as the
nearest neighbour anti-ferro-magnetic order, and can be rationalized as an effect of on-site Coulomb repulsion 
between electrons with opposite spins, is naturally expected to contest $\pi$-conjugation, although modestly.
Disparity between sub-lattices due to zigzag edges, which is well known to consolidate 
nearest neighbour anti-ferro-magnetism in GF, is thus expected to 
considerably influence properties of GNs with substitution by N at N-Nn sites. 

Through unambiguous quantitative estimates of spin resolved bond orders and atomic charges obtained 
based on spatially localized Wannier functions(WF), we find the onset of mechanical strain and loss of
charge-neutrality of N and C atoms due to evolution of the C-N bond-orders in attempt to maximally 
retain  $\pi$-conjugation among C atoms upon substitution by N, to not only govern the energetics of 
substitution but also to favour passivation of N coordinated C atoms as an overall remedy.
Such activated C atoms facilitate covalent adsorption of radicals, di-radicals like O$_2$, 
as well as similarly activated C atoms leading to cross-linking of N doped GNs, although
remain inert to ordinary graphenic C atoms, which is advantageous since it would allow
spatially resolved self-assembly of  N doped GNs. 
Adsorption of O$_2$ as diradical leaves the mono-coordinated O atom charged and reactive
as required for ORR. Activation is found to increase sharply with increasing coordination to N
and proximity to zigzag edges. Absolute magnetization is observed to alter significantly in 
events of adsorption, since it restores $\pi$-conjugation to a variable degree, which in turn
modifies  nearest neighbour anti-ferromagnetism, given the anti-correlation between the two.
In addition to proposing a new paradigm of activation of graphenic C,
this work also highlights the effectiveness of WFs in shaping precise orbital resolved understanding of 
mechanisms of chemical processes.  
\begin{figure}[t]
\centering
\includegraphics[scale=0.35]{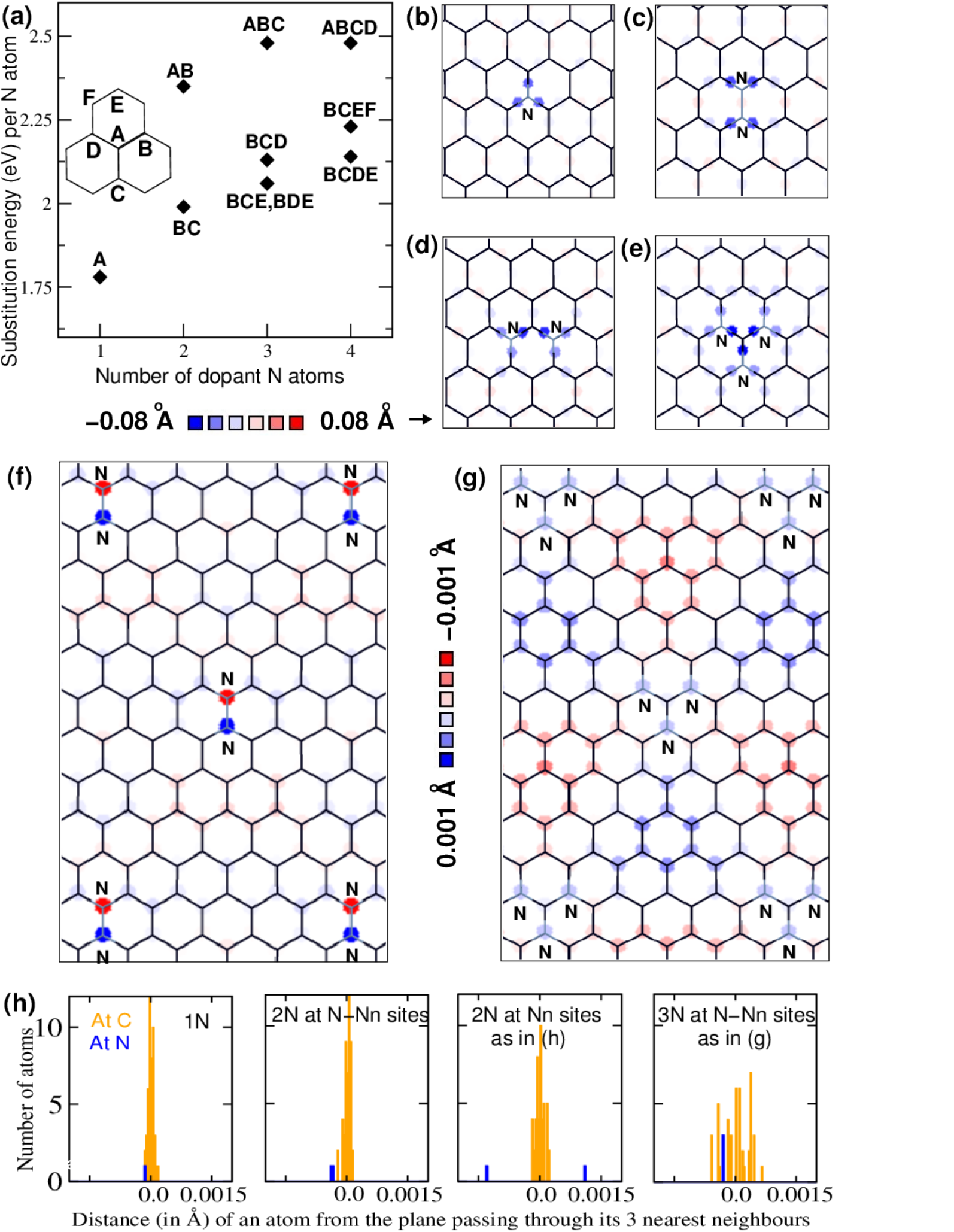}
\caption{(a) Energetics of substitutional doping. 
(b-e) Variation in bond lengths compared to corresponding equilibrium values.
(f,g) Contour plot of the degree of non-co-planarity measured as the vertical distance 
of an atom from the plane passing through its nearest neighbours.
(h) Histogram of the degree of non-co-planarity for different substitution configurations.}
\label{fig1}
\end{figure}

Equilibrium configurations and relevant energetics of substitutional doping are obtained  
within the non-empirical framework of density functional theory (DFT) based on local mean-field
approximation of the many-electron exchange-correlation contributions to total energy. We use the
Quantum Espresso\cite{pwscf} code, which expands wave-functions in the basis of plane waves and
allow  \textit{ultrasoft}\cite{upf} pseudopotentials for valence electrons.
For exchange-correlation we use a gradient corrected Perdew-Burke-Ernzerhof (PBE)\cite{pbe} 
functional. We considered a 5$\times$5 GF super-cell to understand the energetics of substitution by N 
and the proposed activation of C atoms thereby. To study the effect of zigzag edges and use an available 
scheme for unbiased partitioning of total charge density in isolated systems, we also consider an
isolated GF segment(Fig.\ref{fig4}(a)) made of 84 atoms. 
Total energies and configurations are converged with plane-wave cutoffs over 800 eV, $k$-mesh up to 
7$\times$7$\times$1 for the super-cell, and forces less than $10^{-4}$ Rydberg/Bohr using the 
Broyden-Fletcher-Goldfarb-Shanno (BFGS)\cite{bfgs} scheme for total energy minimization. 
To account for the attractive dispersion interactions, semi-empirical 
Grimme approximation\cite{grimme} is used only in cases where chemisorption is not conclusive with PBE.  

Unambiguous  estimates of bond-orders, atomic charges and sub-shell filling, are obtained through 
construction of localized WFs based on joint-diagonalization of 
non-commuting first moment matrices (FMM). 
In one dimension, WFs with maximum localization\cite{jbwf1} 
are naturally the eigen-functions of the FMM which is essentially the 
position operator expanded in the occupied subspace. For periodic systems the FMM is obtained
in terms of geometric phases of Bloch states evolving over the full Brillouin zone.
For isolated systems the FMMs can be directly calculated from the occupied Kohn-Sham (KS) 
eigen-states $\left\{\phi^{KS}_m\right\}$ as 
\begin{equation}
X_{mn}=\langle\phi^{KS}_m|\hat{x}|\phi^{KS}_n\rangle,
\end{equation}
for  occupied KS states in the valence band. 
Diagonalizing a FMM (say $X$ as evaluated above) yields WFs with maximum localization in the
corresponding ($x$) direction. However, owing to the inherent non-commuting nature of the position
operators expanded
within the finite occupied subspace, it is not possible to obtain WFs with maximum localization 
simultaneously in all three directions. Nevertheless, approximate joint diagonalization of 
the three FMMs renders a set of highly localized WFs, which, even 
though strictly not unique,  constitute an unambiguous orbital representation of the occupied sub-space. 
Their non-uniqueness arise only from the sequence in which the three FMMs are considered 
in course of the iterative approximate joint diagonalization process, and in effect has negligible bearing
on the WFs rendered.
Wannier centres (WCs), which are centre of masses of WFs, are readily available as the 
approximate eigen-values of the three FMMs, without having to explicitly 
construct the WFs. 
WCs, each representing one electron for each spin, provide a 
unique dot structure map for valence electrons through out the system. 
In case of partially occupied bands, the map is obtained by weighted sum of WCs for 
different number of KS states:
\begin{equation}
WC(\vec{r})=\sum^{\infty}_{N=1} \left[\sum^N_{i=1} \delta(\vec{\gamma}_i-\vec{r})\right](f_{N+1}-f_N) 
\end{equation}
where $f_N$ is the occupancy of the $N$-th KS state, such that 
\begin{equation}
\int^{\infty}_{-infty} WC(\vec{r}) d\vec{r} = N_e
\end{equation}
where $N_e$ is the total number of valence electrons in the isolated system. 

WCs can be identified in two categories -
(1) associated with atoms, and (2) representing bonds.
Single and double bonds are represented respectively by one and two WCs between two atoms.
For visualization purpose, WCs in close proximity, like the ones representing double bonds, have been 
fused into one WC representing two electrons. WCs are represented by spheres, whose radii are 
scaled linearly by the  number of electrons they represent, which is thus between 0 to 2.
The depth of shading(in gray-scale) of the planar projection of these spheres are also linearly scaled
in the same range. 
Number of electrons associated with an atom is implied by the WCs exclusively associated with it, representing 
unpaired and lone-paired electron(s) if any, and half the order of all the bonds made by the atom 
according to the WCs representing covalent bonds.  
WFs and WCs are constructed using an in-house implementation which takes the KS eigen
states as input. The reason to use this method is that it does not depend on any reference template of orbitals, 
which is required for construction of localised WFs\cite{mlwf,jbwf2} in periodic systems, where it also
biases the resultant WFs towards the set of orbitals specified in the template. 
However, as the method can not be directly extended to periodic systems, estimation of
bond-orders and atomic charges are made based on WCs calculated in hydrogen passivated isolated GF segments,
and the implied mechanisms are argued to be plausible in their corresponding periodic super-cells 
based on the similarities in  bond-lengths, bond-angles and the degrees of 
non-co-planarity of atoms observed in the periodic super-cells and their isolated GF counterparts.

\begin{figure*}[t]
\centering
\includegraphics[scale=0.31]{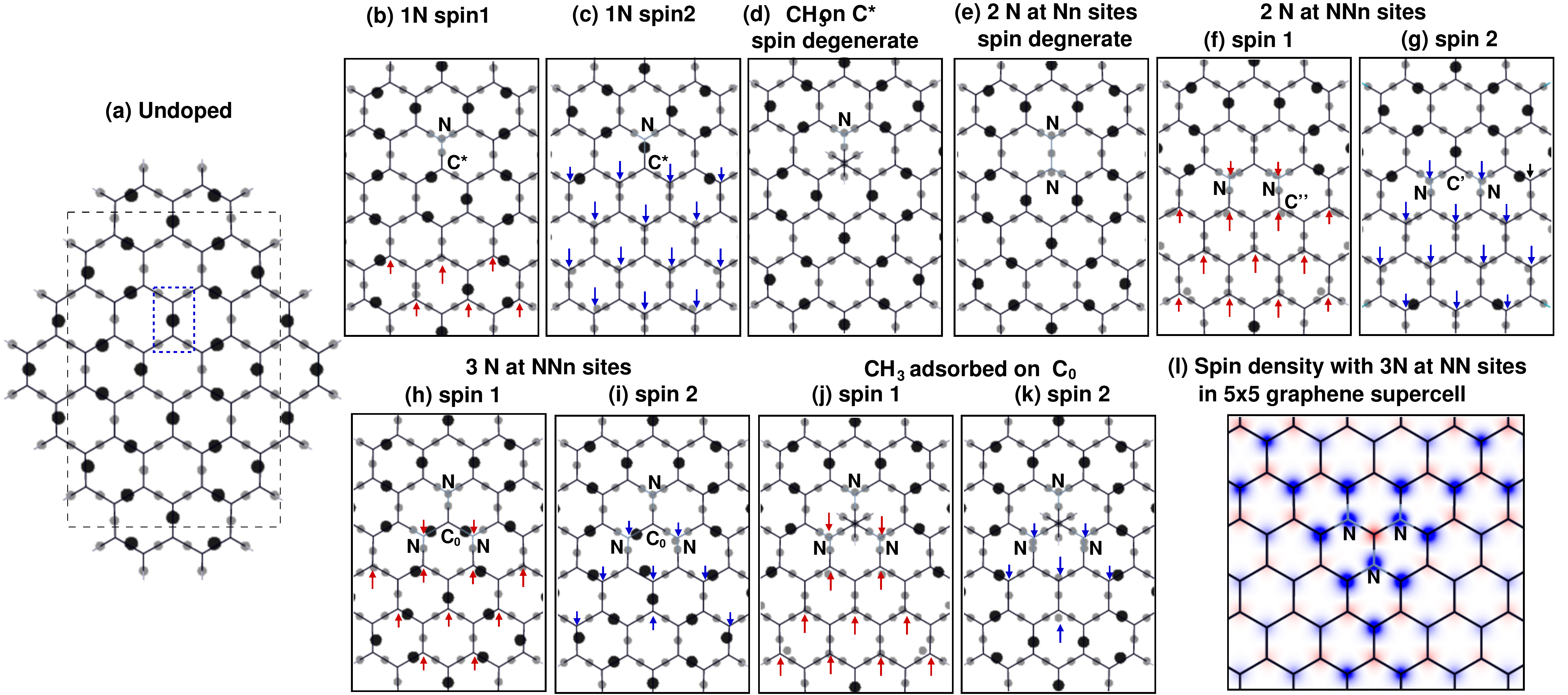}
\caption{(a-k) Planar projection of WCs implying order of bonds, charge state, 
spin polarization  and subshell-filling of atoms. Adjacent WCs have been merged 
into one WC located at their center of mass. Larger and darker circles 
each denote 2 electrons per spin, while the smaller gray circles represent 1 electron per spin.
(l) Spin density ($\rho_{spin1}-\rho_{spin2}$) with substitution at three mutually N-Nn sites
enclosing a carbon atom. }
\label{fig2}
\end{figure*}
Energy required for substitution by N in the periodic super-cell, calculated as 
$(E_{doped}- E_{undopped})+ m E_C- m E_N)$, where $E$ is total energy and $m$ is the number of C atoms substituted,  
is plotted in Fig.\ref{fig1}(a). Consistent with experimental observations \cite{zhao2011,ruitaro2012-2},
Fig.\ref{fig1}(a) implies substitution at nearest neighbouring (Nn) sites to be particularly unfavourable.
The deviation in equilibrium bond-lengths plotted in Fig.\ref{fig1}(b-e) from that of the resonating 
graphenic C-C bonds and single (order 1) C-N and N-N bonds obtained with same parameters, 
suggests strain in the C-N bonds to be possibly responsible for difference in energetics of substitution. 
To translate these deviations into strain, bond orders have to be estimated, since reduction
in length of a bond can also happen with increase in its order.
Fig.\ref{fig2}(e)  suggests complete retention of $\pi$-conjugation
among the C atoms with substitution at Nn sites supporting C-N and N-N bonds or order 1, 
as readily understood by considering substitution at the two sites enclosed by the dashed 
rectangle in Fig.\ref{fig2}(a). The C atoms would thus
strongly favour co-planarity among themselves, while the N atoms, owing to their lone pairs, would strongly 
favour non-co-planarity with their nearest neighbours. We indeed observed this in the periodic super-cell,
as evident in Fig.\ref{fig1}(f), which maps the degree of non-coplanarity of an atom as its vertical distance 
from the plane passing through its three nearest neighbours. 
Thus  with substitution at Nn sites, the C-N bonds in the periodic supercell are also likely to be of order 1, 
implying that those C-N bonds are indeed longitudinally strained as their observed reduction in length
is not due to any increase in their order. 
The stark contrast between C and N atoms in their preference for non-coplanarity with their nearest neighbours 
also suggests non-nominal dihedral strain.

With substitution at non-Nn sites, completely unaltered retention of $\pi$-conjugation
is impossible, resulting into ease in accommodation of C-N single bonds, which are typically 
about 4\% longer than the graphenic C-C bonds. This is evident in Fig.\ref{fig1}(d) which shows 
only two of the six C-N bonds to have significant reduction in length upon bi-substitution at N-Nn sites.  
Furthermore, Fig.\ref{fig2}(c,h,i) suggests increase in order of C-N bonds  with substitution at non-Nn sites, 
despite the consequent loss of charge neutrality of N atoms due to depletion from their lone pairs.
This in effect imply that retention of $\pi$-conjugation among C atoms is the dominant mechanism
which can force locally unfavourable transfer of charges to make way for global lowering of total energy.  
Depletion from lone pairs of N would also imply reduced preference for non-coplanarity of N atoms
with their nearest neighbours, which is indeed evident from Fig.\ref{fig1}(h) as it suggests lesser degree 
of non-coplanarity of N atoms in periodic supercells with non-Nn substitution. 
The corresponding C-N bonds in the periodic supercell are thus expected to be of order > 1 
which would account for their reduced length (Fig.\ref{fig1}(b,d,e)), implying that they are effectively 
unstrained and thus energetically favourable.

\begin{figure}[t]
\centering
\includegraphics[scale=0.33]{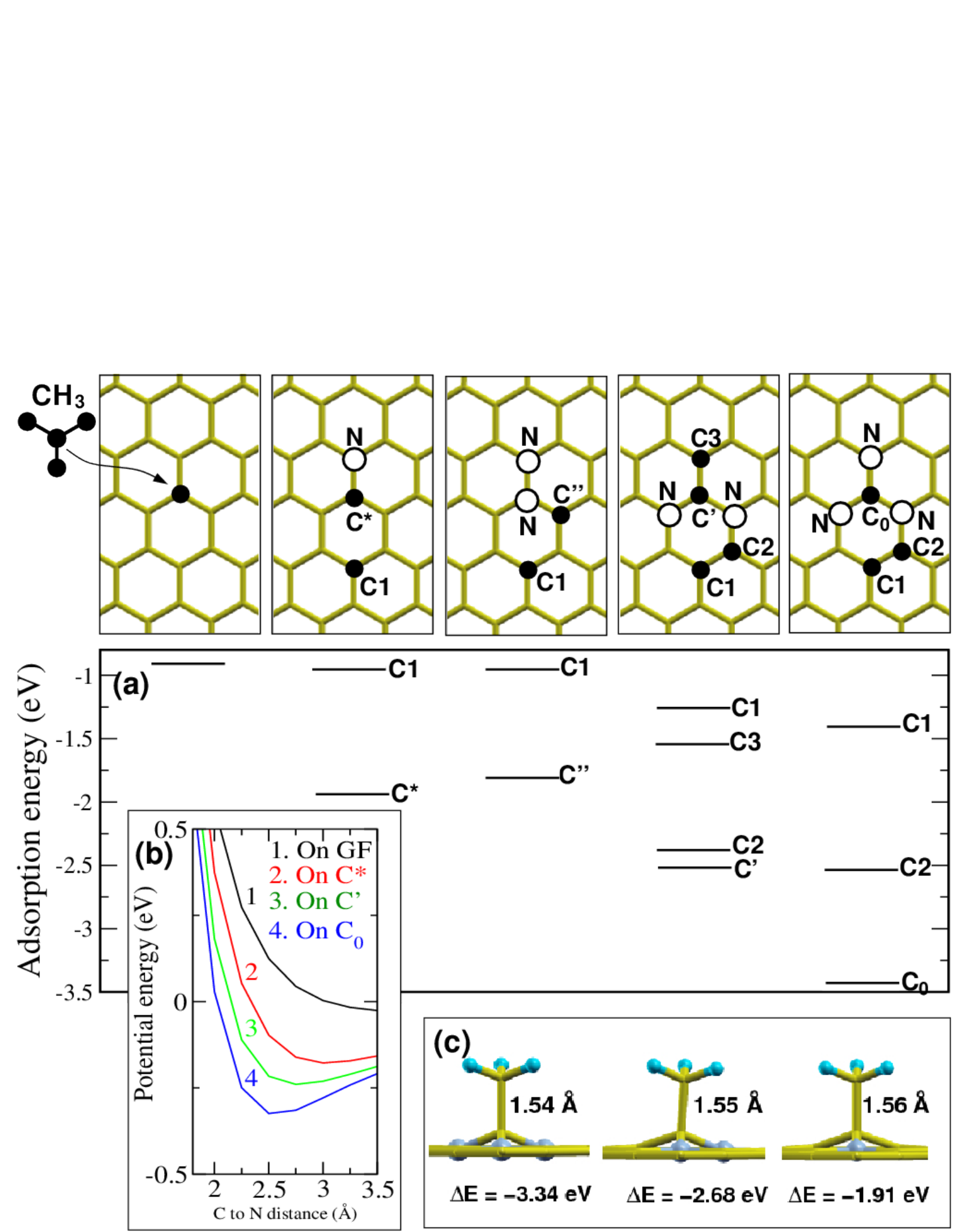}
\caption{(a) Energetics of adsorption of CH$_3$ on different C atoms. 
(b) Interaction potential of CH$_3$ as a function of vertical distance of its C atom
from the C atoms in periodic GF supercells specified in the figure.
(c) Equilibrium configurations of adsorbed CH$_3$ on C$_0$, C' and C* in periodic GF supercell.}
\label{fig3}
\end{figure}

\begin{figure}[t]
\centering
\includegraphics[scale=0.43]{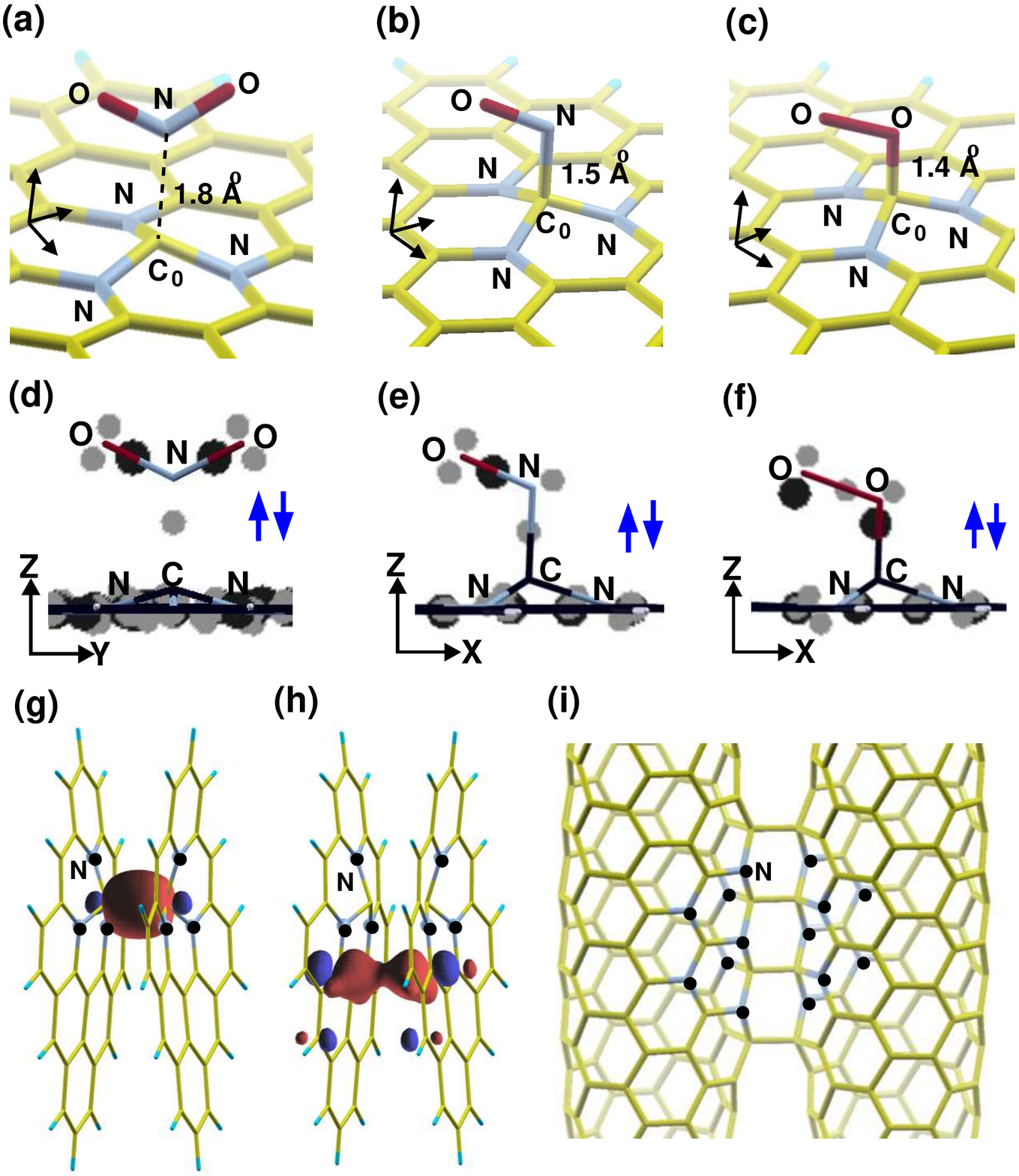}
\caption{Adsorption of NO$_2$ (a), NO (b) and O$_2$ (c) on C$_0$. (d-e) WCs corresponding to
(a),(b) and (c) respectively, depicted by smaller(larger) gray(black) circles denoting
1(2) electrons per spin. (g-h) WFs representing the C$_0$-C$_0$ $\sigma$-bond (g) and an
inter-segement $\pi$-$\pi$ bond (h). (i) Inter-CNT C$_0$-C$_0$ and adjoining
C-C bonds. N atoms are marked by black dots in (g-i). }
\label{fig4}
\end{figure}

Hindered $\pi$-conjugation is expected to allow the nearest neighbour
anti-ferromagnetic correlation to consolidate, which is further expected to reinforce
due to electron localization induced by the zigzag edges in the vicinity.
In the isolated segment spin-separation is observed (Fig.\ref{fig2}(b-c,f-i)) 
with all possible non-Nn substitution, whereas in the periodic supercell spin separation consolidates 
(Fig.\ref{fig2}(l)) only with tri-substitution at mutually N-Nn sites enclosing a common nearest 
neighbouring C atom(C$^*$ in Fig.\ref{fig3}). This again reiterates that the C atoms would always like 
to maximize $\pi$-conjugation among themselves, which can be resisted effectively 
by the nearest neighbour anti-ferromagntic order only when all the three 
$\pi$-conjugation configurations are equally compromised, as happens with enclosing tri-substitution, 
or in the presence of magnetic impurity or zigzag edges.
Within the region of spin-separation each C atom has in effect an unpaired electron centred on 
them (Fig.\ref{fig2}(b-c,f-i)), indicating their partial $sp^3$ hybridization, which is 
expected to help cooperatively in allowing the local non-coplanarity preferred by the N atoms. 
However, as the  N atoms themselves have reduced preference for non-coplanarity due to depletion
from their lone pairs, an effectively planar equilibrium configuration is attained with mild systematic 
corrugation due to the partial $sp^3$ hybridization of anti-ferromagnetically correlated C atoms, 
as evident in Fig.\ref{fig1}(g) for the periodic supercell with enclosing tri-substitution. 

Energetics of adsorption of Me (CH$_3$) radical on C atoms shown in Fig.\ref{fig3}(a) 
suggests adsorption on the N coordinated C atom (C$_0$) due to enclosing tri-substitution 
to be prominently most favourable. 
As evident from Fig.\ref{fig2}(b,c), mono-substitution by N in the isolated segment leads to spin-separation
and a C-N double bond for the majority spin.  Interestingly, passivation of a C (C$^*$ in Fig.\ref{fig3}) 
atom nearest to the N atom, exemplified here through adsorption of Me, not only  
restores complete $\pi$-conjugation by quenching spin-separation, but also reduces the order of 
the C-N double bond (Fig.\ref{fig2}(c)) to 1, which restores charge neutrality of the N atom and 
relieves the tensile stresses in the adjacent graphenic C-C bonds. 
In case of substitution at Nn sites, which retains $\pi$ conjugation among C atoms intact, 
passivation of any C atom beyond the nearest neighbourhood of N atoms is same as passivation of a 
C atom in an undoped segment as evident in Fig.\ref{fig3}(a).
Passivation of C atoms (C" in Fig.\ref{fig3}) next to an N atom is more favourable on account 
of cooperative non-planarity of the N and the passivated C atom, although it hinders $\pi$ conjugation locally
and consolidates nearest-neighbour antiferromagnetic order in the vicinity.
With the experimentally observed substitution at N-Nn sites, 
the C atom  (C' in Fig.\ref{fig3}) nearest to both the N atoms, neither retains 
charge neutrality nor completes sub-shell filling. Interestingly, both the shortcomings
are restored upon passivation of C', which also makes the C atom next to it charge neutral, 
implying a higher level of activation of C' than an ordinary graphenic C atom, as also evident 
in Fig.\ref{fig3}(a).

With enclosing tri-substitution, order of C$_0$-N bonds (Fig.\ref{fig2}(h,i)) 
increases to around 1.5 on the average, implied by the three C-N double bonds
out of six total C-N bonds accounting for the two spins separately.  
This is consistent with the shortest C-N bond lengths observed in the periodic super-cell among the
different possible N substitution configurations, implying high degree of mechanical stress
in the adjacent C-C bonds.
While strong depletion from lone-pairs of N atoms leaves them positively charged and helps the nearest 
neighbour anti-ferromagnetism to consolidate, 
C$_0$ would have excess charge centered on it owing to the increased order of the C$_0$-N bonds.
This excess charge on C$_0$, being in excess to what it requires for complete sub-shell feeling, would
increase total energy and should have low ionization potential. 
Conveniently, as evident in (Fig.\ref{fig2}(j,k)), complete $sp^3$ hybridization of C$_0$, 
as possible upon passivation, 
would force all C$_0$-N single bonds to be of order 1, which will restore charge neutrality of 
of N atoms, besides reducing tensile stresses in the surrounding C-C bonds and facilitating 
cooperative non-planarity of C$_0$ and N atoms. 
C$_0$ would thus be prone to passivation, and thereby, non-trivially active in terms
of adsorption of radicals, as evident from the adsorption energies plotted in Fig.\ref{fig3}(a). 
Interaction potential $(E_{(Me+GF)}-E_{Me}-E_{GF})$ between Me and N doped GF plotted in  Fig.\ref{fig3}(b)
suggests consistent increase in the level of activation of C atoms with their increasing coordination
to N, which is directly corroborated by the adsorption energies quoted in Fig.\ref{fig3}(c).
Spin-separation reduces significantly upon passivation of C$_0$ in the periodic supercell primarily 
due to restoration of full occupancy of the lone-pair orbitals of N.

To generalize the activation of C$_0$ suggested by covalent adsorption of Me, 
we study adsorption of two other hazardous free radicals - NO and NO$_2$.  
As evident in Fig.\ref{fig4}(a,b,d,e), adsorption of NO and NO$_2$  both happens through 
C$_0$-N single bonds. Unlike in isolated NO, the N atom in adsorbed NO completes sub-shell filling as 
well as charge neutrality(Fig.\ref{fig4}(e)). 
In adsorbed NO$_2$, both N and O atoms are charge neutral, although the N atom is over coordinated on
account of the two N-O double bonds suggested by the WCs, in agreement with  N-O bond lengths of about 1.2\AA.
Notably, an isolated NO$_2$ molecule has resonating double and single N-O bonds with 
the N (O) atom(s)  positively (negatively) charged with incomplete (complete) sub-shell filling. 
The adsorbed configuration with two N-O double bonds is stable since the energy of the electron 
at N in excess to what it requires to complete its sub-shell filling, is lowered due to its involvement
 in the C$_0$-N covalent bond. The over coordination of N is reflected in the long
length (1.8\AA) of the  C$_0$-N bond. 

\begin{figure*}[t]
\centering
\includegraphics[scale=0.43]{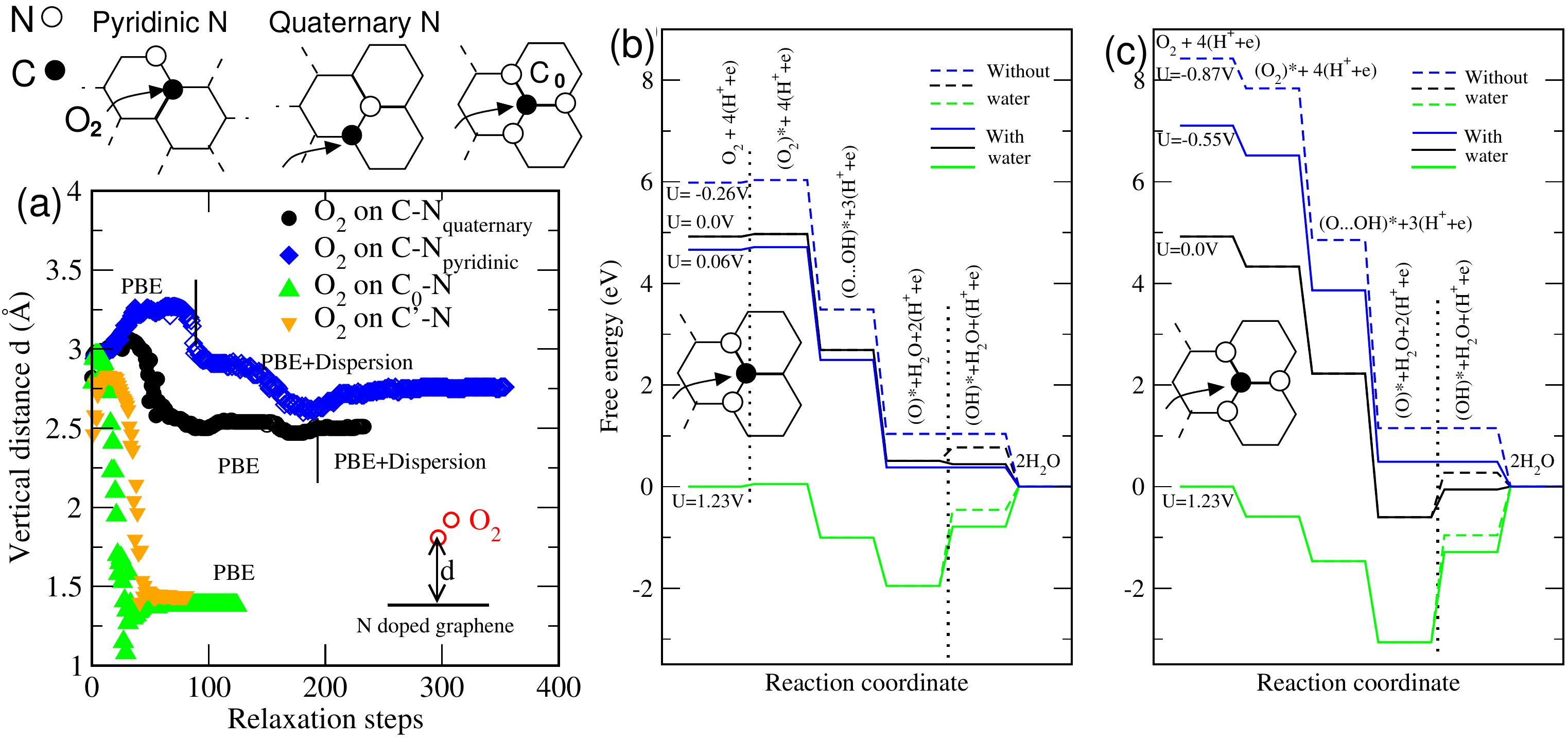}
\caption{(a) Distance between the lower O of O$_2$, and the active C atoms as shown in the upper panel,
in course of structural relaxation using the BFGS\cite{bfgs} scheme. DFT + Dispersion indicates use of 
Grimme's approximation\cite{grimme} along with PBE exchange-correlation to account for the dispersion interation.
(b-c) Free energy diagram for complete reduction of the diatomic oxygen adsorbed on the active C atoms
shown in the inset. These free diagrams are to be compared with free diagram\cite{studt} for ORR on a C atom
next to a quaternary N atom, which shows an overpotential of about 0.7 V.}
\label{fig5}
\end{figure*}
Encouraged by favourable adsorption of radicals on C$_0$, we now probe in relation to  ORR catalysis, the 
adsorption of O$_2$, which is known to exist in atmosphere as diradical. 
Fig.\ref{fig5}(a) unambiguously suggests  spontaneous covalent adsorption of diatomic O 
on C$_0$ and C', whereas, non-nominal activation barriers appear to exist for adsorption of
diatomic O on C atoms next to quaternary and pyridinic N atoms. Notably, pyridinic N atoms at graphene edges
and quaternary N atoms embedded in graphene are known to adsorb oxygen through charge transfer.
Ground state of O$_2$ is a triplet where the two O atoms, each having an unpaired electron,
are connected by an O-O single bond.  
As per WC distribution, O$_2$ appears (Fig.\ref{fig4}(f)) to be adsorb by C$_0$ through a O-C$_0$ 
double bond.
In adsorbed O$_2$ both the O atoms complete sub-shell filling unlike in the O$_2$
diradical, but both become charged in the process. 
The mono-coordinated outer O atom clearly has an extra electron (Fig.\ref{fig4}(f)), while the O atom
coordinated to C$_0$ has an electron less. Thus the O$_2$ on the whole remain charge neutral, although
their host C$_0$ acquires an extra electron from the rest of the GN to support its over-coordination 
upon adsorption of  O$_2$. The +1e charge of the O atom coordinated C$_0$ justifies the longer length 
(1.38\AA) of the O-C$_0$ double bond.
The negatively charged outer O atom, although would attract  cations, but may not be immediately able 
to donate its excess charge since such a donation would render its sub-shell filling incomplete. 
However, it can adsorb a radical and complete its sub-shell filling after donating its 
excess charge to a cation, or to the host GN itself which lost 1e to C$_0$
upon adsorption of O$_2$. 
Interestingly, the O-O bond  dissociates spontaneously upon adsorption of H, which can be crucial 
for facilitation of ORR since the outer O atom thus acts as a free active atomic O. 

Next we study the effectiveness of the experimentally observed substitution configurations to
ORR through  construction of free energy diagrams\cite{norskov1,norskov2} for complete reduction of an 
adsorbed  diatomic oxygen on C' and C$_0$ (Fig.\ref{fig3}). Free energy of a reduction step as a 
function of an applied voltage U 
is calculated as $\Delta G(U) = G(0) + neU - 4G(H_2O)$\cite{norskov1}, where G is the Gibbs free energy 
calculated as $G=E+ZPE-TS$,  $n$ is the number of atomic H (proton+electron)
per $O_2$,  $e$ bring the electronic charge, and ZPE the zero point energy calculated from the vibrational 
frequecies. 
Notably, in the free energy calculations the H atoms are considered to be at equilibrium with molecular H$_2$.
The entropy S is considered only for molecules in gas phase and has been taken from
standard reference\cite{stand}.
Free energy diagram for reduction of O$_2$ on C atoms next
to isolated quaternary N atoms suggests\cite{studt} an overpotential potential of about 0.7V which is consistent
with observation of ORR with N-doped graphene in alkaline media. However, lack of adsorption of
diatomic O on such C atoms can pose a limitation in terms of the availability of O for ORR.
C' or C$_0$ atoms on the other hand, facilitate adsorption of  diatomic O on those C atoms as a diradical,
but tend to bind atomic O strongly enough, more than OH, to make the third reduction step, as
described in Fig.\ref{fig5}(b,c), uphill in  free energy for U=0V.
ORR on those C atoms thus can only be completed through application of a large overpotential to make all the 
reduction steps downhill (blue lines in Fig.\ref{fig5}(b,c)), which is undesirably more than the equilibrium 
potential of U=1.23V. 
Thus the thermodynamically limiting step for ORR on C' or C$_0$ is the reduction of adsorbed O to OH, 
whereas for the C atoms next to the quaternary N atoms the rate limiting step is the reduction of the 
adsorbed OOH or OH\cite{studt}, which is also typically the case on metal surfaces\cite{norskov2}. 
Incorporating extra stabilization\cite{norskov1} 
of adsorbed OH due to hydrogen bonding provided by the water molecules in the vicinity, the overpotental 
requirement reduces to a value marginally less than the equilibrium potential. Since increase in pH
from zero has same effect ($G(pH)=KTln(10)pH$\cite{norskov1}, $K$ being the Boltzmann constant) as that of 
lowering of an applied voltage (U), high overpotential implies ORR to be
possible only in an acidic media, as has been reported\cite{orracid1,orracid2}.
To summarize the discussion, in one hand, next to the quaternary N atoms, we have the C atoms which appears 
to support ORR in acidic as well as alkaline media, but may be limited by the availability of O itself due to
poor adsorption of diatomic O which dissociates only upon adsorption followed by reduction. On the hand we have 
the C' and C$_0$ atoms, which are coordinated to more than one N atoms, and ensure availability of O by 
supporting spontaneous adsorption of diatomic O, but
having high overpotential, which stipulates an acidic media. We recall here that, experimentally, N atoms have 
been observed to substitute at next nearest sites which support existence C' and C$_0$, which has been
also found to be energetically favourable. Thus it appears to be a balance between the
availability of O facilitated through adequate adsorption of O$_2$, and the ease in their complete reduction in terms 
of lowering of free energy, is what would finally determine the effectiveness of N-doped graphene as a 
good metal-free alternate for ORR catalysis. 

In relation to mutual covalent adsorption of N doped GNs, we now probe whether C$_0$ can 
be passivated by a similar C atom in another N doped GN. 
As already shown, $sp^3$-hybridization of C$_0$ due to its passivation, would 
preserve planarity of the rest of the GF. Steric repulsion between two such planar parallel 
GF segments is expected to make their mutual adsorption through inter-GF C$_0$-C$_0$ covalent bonds 
difficult due to proximity.
However, additional interactions such as the dispersion interaction(D) and $\pi-\pi$ stacking interaction 
are expected to help in inter-GF cohesion.
Strength of these interactions dependent on the nature of stacking of the segments.
We study cohesion between two isolated segments starting from an initial configuration
where they are 3 \AA apart. Steric repulsion as well as  $\pi-\pi$ stacking interaction both are expected to 
intensify with AA stacking. Interestingly, covalent adsorption between N-doped GF segments 
due to inter-GF C$_0$-C$_0$ covalent bond (Fig.\ref{fig4}(g)) is observed only with AA stacking,
even without the support of dispersion interaction.
Thus with AA stacking the inter-segment $\pi-\pi$ interaction, as represented by the WF shown in 
Fig.\ref{fig4}(h), is stronger than the inter-segment steric repulsion. No inter-GF adsorption with AB 
stacking implies sharper drop in  $\pi-\pi$ interaction than that in steric repulsion compared to 
AA stacking, such that, the dispersion interaction\cite{grimme} alone can not overcome 
the potential barrier induced by steric repulsion. 
Notably, as C$_0$ can not be passivated by an ordinary graphenic C atom, N doping can be
used to define regions of chemisorption between GF segments. 

However, with reduced steric repulsion, cohesion between N doped CNTs appears to be easier and generic. 
As evident in Fig.\ref{fig4}(i), unlike in case of planar segments, additional inter-CNT C-C bonds are formed
in the neighbourhood of the inter-CNT C$_0$-C$_0$ bonds, implying enhanced activation of C atoms in the 
neighbourhood of C$_0$ in CNT. This is straightforward since the contractile strain caused by the 
C$_0$-N bonds of higher order causes reduction in surface area, which in turn increases local curvature,
and thereby, the level of $sp^3$-hybridization as well, leading to an enhanced level of chemical 
activation of C atoms in the neighbourhood of substitution by N in CNT than that in GF.  

To conclude, we reveal from first principles a general scenario of non-trivial chemical activation of 
N coordinated graphenic C atoms upon substitution at  next nearest sites as experimentally observed. 
Mechanical strain due to increased order of C-N bonds and loss of charge neutrality of N and C atoms,
in attempt to maximally retain $\pi$-conjugation among C atoms, is central to the energetics 
of substitution by N, as well as the mechanisms leading to activation of N coordinated C atoms.
Substitution at nearest neighbouring sites is found to be unfavourable on account of high
mechanical stresses stemmed at the non-co-planarity preferred by the N atoms amidst strongly co-planer
C atoms due to complete retention of $\pi$-conjugation among them.
Activation is most pronounced for C atoms with all three nearest neighbours substituted
by N possible upon tri-substitution at mutually N-Nn sites.
Proximity to zigzag edges and curvature naturally enhances the level of activation of N coordinated C atoms. 
Relevant to oxygen reduction reactions (ORR), adsorption of diradical O$_2$ on 
activated C atoms coordinated to more than one N atoms, renders the mono-coordinated outer O atom reactive, 
although the overpotential requirement for complete reduction of an adsorbed diatomic O is high due to 
strong binding of O on such C atoms and should allow only an acidic media.
Between two adjacent GFs or CNTs, mutual covalent adsorption exclusively among the N coordinated 
C atoms and possibly few more C atoms in their close vicinity, promises a new
route to robust and ordered self-assembly of GNs.

Calculations were performed partly using the central facility for high performance computing at 
NISER supported by the Dept. of Atomic Energy of the Govt. of India (GOI), and partly in another facility 
supported by a financial grant (SR/NM/NS-1026/2011) from the Dept. of Sci. and Tech. of GOI.  
 

\begin{thebibliography}{99}
\bibitem{yu2010}
Yu, D.; Nagelli, E,; Du, F.; Dai, L.
Metal-free Carbon Nanomaterials Become More Active
than Metal Catalysts and Last Longer.
{\it J. Phys. Chem. Lett.} {\bf 2010}, {\it 1}, 2165-2173.
\bibitem{wang2012} 
Wang, H.; Maiyalagan, T.; Wang, X.
Review on Recent Progress in Nitrogen-Doped Graphene: Synthesis,
Characterization, and its Potential Applications.
{\it ACS Catal.} {\bf 2012},  {\it 2}, 781-794.
\bibitem{ruitaro2012} Lv, R.; Terrones, M.
Towards New Graphene Materials: Doped Graphene Sheets and Nanoribbons.
{\it Mat. Let.} {\bf 2012}, {\it 78}, 209-218.
\bibitem{gong2009} 
Gong, K.; Du, F.; Xia, Z.; Durstock, M.; Dai, L. Nitrogen-doped
Carbon Nanotube Arrays with High Electrocatalytic Activity for
Oxygen Reduction.{\it Science} {\bf 2009},  {\it 323}, 760−764.
\bibitem{qu2010}  Qu, L.; Liu, Y.; Baek, JB; Dai, L.
Nitrogen-doped Graphene as Efficient Metal-free Electrocatalyst for 
Oxygen Reduction in Fuel Cells.{\it ACS Nano}  {\bf 2010},  {\it 4(3)}, 1321-1326.
\bibitem{jafri2010} Jafri, RI; Rajalakshmi, N.; Ramaprabhu, S.
Nitrogen doped Graphene Nanoplatelets as Catalyst Support 
for Oxygen Reduction Reaction in Proton Exchange Membrane Fuel Cell.
{\it J. Mater. Chem.} {\bf 2010},  {\it 20}, 7114-7117.
\bibitem{sheng2011} Sheng, Z-H; Shao, L; Chen, J-J; Bao, W-J; Wang, F-B;  Xia, X-H.
Catalyst-free Synthesis of Nitrogen-doped Graphene via Thermal Annealing Graphite 
Oxide with Melamine and its Excellent Electrocatalysis.
{\it ACS Nano} {\bf 2011},  {\it 5}, 4350-4358.
\bibitem{li2012}  Li, Y.;  Zhou, W.;  Wang, H.; Xie, L.;  Liang, Y.; Wei, F.; 
Idrobo, J-C;  Pennycook, S-J; Dai, H.
An Oxygen Reduction Electrocatalyst Based on Carbon Nanotube-Graphene Complexes.
{\it Nat. Nano.} {\bf 2012}, {\it 7}, 394-400.
\bibitem{sharifi2012} Sharifi, T.; Hu, G.; Jia, X.; Wågberg, T. 
Formation of Active Sites for Oxygen Reduction Reactions by Transformation of 
Nitrogen functionalities in Nitrogen-doped Carbon Nanotubes.
{\it ACS Nano} {\bf  2012},  {\it 6(10}), 8904-8912.
\bibitem{cheng2012} Chen, S.; Bi, J.; Zhao, Y.; Yang, L.; Zhang, C.;
 Ma, Y; Wu, Q.; Wang, X.;  Hu, Z.
Nitrogen‐Doped Carbon Nanocages as Efficient Metal‐Free Electrocatalysts for 
Oxygen Reduction Reaction. {\it  Adv. Mater.} {\bf 2012},  {\it 24}, 5593-5597.
\bibitem{zhao2013} Zhao, y.; Yang, L.; Chen, S.; Wang, X.; Ma, Y.; 
Wu, Q.;  Jiang, Y.;  Qian, W.;  Hu, Z.
Can Boron and Nitrogen Co-doping Improve Oxygen Reduction Reaction Activity of 
Carbon Nanotubes?
{\it J. Am. Chem. Soc.} {\bf 2013},  {\it 135(4)}, 1201-2014.
\bibitem{vikkisk2014} Vikkisk, M.; Kruusenberg, I.;  Joost, U.;  Shulga, E.; 
Kink, I.; Tammeveski, K. Electrocatalytic Oxygen Reduction on 
Nitrogen-doped Graphene in Alkaline Media. 
{\it App. Catal. B: Env.} {\bf 2014},  {\it 147}, 369-376.
\bibitem{chung2013}  Chung, H. T.;  Won, J. H.; Zelenay, P. 
Active and Stable Carbon Nanotube/Nanoparticle Composite 
Electrocatalyst for Oxygen Reduction.
{\it Nat. Comm.} {\bf 2013} {\it 4} 1922(1-5).
\bibitem{chen2013} Chen, P.;  Xiao, T. Y.;  Qian, Y. H.; Li, S. S.;  Yu, S. H.
A Nitrogen‐Doped Graphene/Carbon Nanotube Nanocomposite with Synergistically 
Enhanced Electrochemical Activity.
{\it Adv. Mat.} {\bf 2013}, 25, 3192-3196.
\bibitem{ratso2014} Ratso, S.; Kruusenberg, I.; Vikkisk, M.; Joost, U.; 
Shulga, E.; Kink, I.; Kallio, T.; Tammeveski, K. 
Highly Active Nitrogen-doped Few-layer Graphene/Carbon Nanotube Composite 
Electrocatalyst for Oxygen Reduction Reaction in Alkaline Media.
{\it CARBON }{\bf 2014}, 73, 361-370.
\bibitem{tian2014}  Tian, G. L.; Zhao, M. Q.; Yu, D.; Kong, X. Y.;  Huang, J. Q.; Zhang, Q.; Wei, F.
Graphene Hybrids: Nitrogen-Doped Graphene/Carbon Nanotube Hybrids: 
In Situ Formation on Bifunctional Catalysts and Their Superior 
Electrocatalytic Activity for Oxygen Evolution/Reduction Reaction.
{\it Small}, {\bf 2014}, 10, 2113.
\bibitem{higgins2014}  Higgins, D.C.; Hoque, Md A.;  Hassan, F.;  Choi, J-Y; Kim, B.;  Chen, Z. 
Oxygen Reduction on Graphene–Carbon Nanotube Composites Doped Sequentially with 
Nitrogen and Sulfur.
{\it ACS Catal.}, {\bf 2014}, 4(8) 2734-2740. (2014).
\bibitem{payne}  Nevidomskyy, A. H.; Gabor, C.; Payne, M. C. 
Chemically Active Substitutional Nitrogen Impurity in Carbon Nanotubes.
{\it Phys. Rev. Lett.} {\bf 2003}, 91, 105502(1-4).
\bibitem{wang2011} Wang, Z.; Jia, R.; Zheng, J.; Zhao, J.; Li, L.; Song, J.; Zhu, Z. 
Nitrogen-promoted Self-assembly of N-doped Carbon Nanotubes and their Intrinsic Catalysis for Oxygen 
Reduction in Fuel cells.
{\it ACS Nano} {\bf 2011}, 5(3), 1677-1684.
\bibitem{hu2010} Hu, X.;  Wu, Y.;  Li, H.;  Zhang, Z. 
Adsorption and Activation of O2 on Nitrogen-doped Carbon Nanotubes
{\it J. Phys. Chem. C} {\bf 2010} 114, 9603-9607. 
\bibitem{kim2011} Kim, H.;  Lee, K.;  Woo, S. I.; Jung, Y. 
On the Mechanism of Enhanced Oxygen Reduction Reaction in Nitrogen-doped Graphene Nanoribbons.
{\it Phys. Chem. Chem. Phys.} {\bf 2011}, 13, 17505-17510.
\bibitem{zhang2011} Zhang, L.; Xia, Z.
Mechanisms of Oxygen Reduction Reaction on Nitrogen-doped Graphene for Fuel-cells.
{\it J. Phys. Chem. C} {\bf 2011}, 115(22), 11170-11176.
\bibitem{yu2011} Yu, L.;  Pan, X.; Cao, X.;  Hu, P.;  Bao, X.
Oxygen Reduction Reaction Mechanism on Nitrogen-doped Graphene: A Density Functional Theory Study
{\it J. Catal.} {\bf 2011 }, 282, 183-190.
\bibitem{yan2012} Yan, H.,J.; Xu, B.; Shi, S., Q.; Ouyang, C.,Y.
First-principles Study of the Oxygen Adsorption and Dissociation on Graphene and Nitrogen doped 
Graphene for Li-air Batteries
{\it J. App. Phys.} {\bf 2012}, 112, 104316(1-5).
\bibitem{zhao2011} Zhao, L.;  He, R.; Taeg Rim, K.; Schiros, T.;  Soo Kim, K,;  Zhou, H.; 
Gutiérrez, C.; Chockalingam, S. P.;  Arguello, C., J.; Pálová, L.; et al. 
Visualizing Individual Nitrogen Dopants in Monolayer Graphene.
{\it Science} {\bf 2011}, 333, 999-1003.
\bibitem{ruitaro2012-2} Lv, R.; Li, Q.; Botello-Me´ndez, A., R.; Hayashi, T.; Wang, B.; Berkdemir, A.; 
Hao, Q.; Elias, A. L.;  Cruz-Silva, R.;  Gutie´rrez, H. R.; et al.  
Nitrogen-doped Graphene: Beyond Single Substitution and Enhanced Molecular Sensing.
{\it Sci. Rep.} {\bf 2012} 2, 586(1-8).
\bibitem{pwscf} Giannozzi, P.; Baroni, S.; Bonini, N.; Calandra, M.; Car, R.; Cavazzoni, C.; Ceresoli, D.; 
Chiarotti, G., L.; Cococcioni, M.; Dabo, I.; et al.  
 QUANTUM ESPRESSO: A Modular and Open-source Software Project for Quantum Simulations of Materials.
{\it  J. Phys.: Cond. Mat.}  {\bf 2009} 21 395502(1-20)
\bibitem{pbe} Perdew, J., P.;  Burke, K.; Ernzerhof, M. 
Generalized Gradient Approximation made simple.
{\it Phys. Rev. Lett.} {\bf 1996}, 77, 3865-3868.
\bibitem{upf} Vanderbilt, D. 
Soft Self-consistent Pseudopotentials in a Generalized Eigenvalue Formalism
{\it Phys. Rev. B} {\bf 1990} 41(R) 7892-7895.
\bibitem{bfgs}Fletcher, R.
{\it Practical Methods of Optimization};
Wiley: New York; 1987.
\bibitem{grimme}Grimme, S. 
Semiempirical GGA‐type Density Functional Constructed with a Long‐range Dispersion Correction
{\it J. Comp. Chem.} {\bf 2006}, 27, 1787-1799.
\bibitem{wan1} Wannier, G. H.;
The Structure of Electronic Excitation Levels in Insulating Crystals.
{\it Phys. Rev.} {\bf 1937}, 52, 191-197. 
\bibitem{wan2} Kohn, W,;
Construction of Wannier Functions and Applications to Energy bands.
{\it Phys. Rev. B}, {\bf 1973}, 7, 4388-4398.
\bibitem{jbwf1}  Bhattacharjee, J.; Waghmare, U., V.; 
Geometric phases and Wannier functions of Bloch electrons in One Dimension.
{\it Phys. Rev. B}  {\bf 2005}, 71, 045106(1-5).
\bibitem{mlwf} Marzari, N.; Vanderbilt, D. 
Maximally Localized Generalized Wannier functions for Composite Energy Bands.
{\it Phys. Rev. B} {\bf 1997}, 56, 12847-12865.
\bibitem{jbwf2} Bhattacharjee, J.;  Waghmare, U., V. 
Localized Orbital Description of Electronic Structures of Extended Periodic Metals, 
Insulators, and Confined Systems: Density Functional Theory Calculations.
{\it Phys. rev. B (R)} {\bf 2006}, 73 121102(1-4).
\bibitem{norskov1} Nørskov, J., K.;  Rossmeisl, J.; Logadottir, A.; Lindqvist, L. 
Origin of the Overpotential for Oxygen Reduction at a Fuel-cell Cathode.
{\it J. Phys. Chem. B} {\bf 2004} 108(46), 17886-17892.
\bibitem{norskov2} Viswanathan, V.;  Hansen, H., A.;  Rossmeisl, J.; Nørskov, J., K.
Universality in Oxygen Reduction Electrocatalysis on Metal Surfaces.
{\it ACS Catal.} {\bf 2012} 2(8), 1654-1660.
\bibitem{stand} Chase Jr., M., W. {\it NIST-JANAF Thermochemical Tables, 4th ed.}
J. Phys. Chem. Ref. Data, Monograph 9, 1998.
\bibitem{studt} Studt, F.
The Oxygen Reduction Reaction on Nitrogen-Doped Graphene.
{\it Catal. Lett.} {\bf 2013} 143, 58-60.
\bibitem{orracid1} Le, K., R.; Lee, K., U.;  Lee, J., W.; Ahn, B., T.; Woo, S., I. 
Electrochemical Oxygen Reduction on Nitrogen Doped Graphene Sheets in Acid Media.
Electrochem. Comm. {\bf 2010}, 12, 8, 1052-1055.
\bibitem{orracid2}Zhang, L.;  Niu, J.; Dai, L.;  Xia, Z.
Effect of Microstructure of Nitrogen-doped Graphene on Oxygen Reduction Activity in Fuel Cells
{\it Langmuir} {\bf 2012}, 28 7542-7550.
\end{thebibliography}
\end{document}